\pdfoutput=1

\documentclass[11pt,twoside,a4paper,cmspaper,final,collab]{cms-tdr}
\def\svnVersion{433300P}\def\cmsCernNoTag{CERN-EP-2017-143}\def\cmsCernDate{\today}\def\cmsMessage{Published in Physics Letters B as \href{http://dx.doi.org/10.1016/j.physletb.2017.10.021}{\doi{10.1016/j.physletb.2017.10.021.}}}

\begin{document}\cmsNoteHeader{HIG-17-011}

\hyphenation{had-ron-i-za-tion}
\hyphenation{cal-or-i-me-ter}
\hyphenation{de-vices}
\RCS$Revision: 399995 $
\RCS$HeadURL: svn+ssh://svn.cern.ch/reps/tdr2/papers/HIG-17-011/trunk/HIG-17-011.tex $
\RCS$Id: HIG-17-011.tex 399995 2017-04-20 17:55:45Z hroskes $
\newlength\cmsFigWidth
\ifthenelse{\boolean{cms@external}}{\setlength\cmsFigWidth{0.85\columnwidth}}{\setlength\cmsFigWidth{0.4\textwidth}}
\ifthenelse{\boolean{cms@external}}{\providecommand{\cmsLeft}{upper\xspace}}{\providecommand{\cmsLeft}{left\xspace}}
\ifthenelse{\boolean{cms@external}}{\providecommand{\cmsRight}{lower\xspace}}{\providecommand{\cmsRight}{right\xspace}}
\providecommand{\CL}{CL\xspace}
\providecommand{\CLend}{CL.\xspace}
\newcommand{\V}{\ensuremath{\cmsSymbolFace{V}}\xspace}
\newcommand{\X}{\ensuremath{\cmsSymbolFace{X}}\xspace}
\newcommand{\qq}{\ensuremath{\PQq\PQq}\xspace}
\newcommand{\ff}{\ensuremath{\mathrm{ff}}\xspace}
\newcommand{\ffbar}{\ensuremath{\cmsSymbolFace{f}\overline{\cmsSymbolFace{f}}}\xspace}
\newcommand{\usedLumiAnofb}{5.1}
\newcommand{\usedLumiB}{19.7\fbinv}
\newcommand{\usedLumiCnofb}{2.7}
\newcommand{\usedLumiD}{35.9\fbinv}
\newcommand{\mllll }{m_{4\ell}}
\newcommand{\mlplm}{m_{\ell\ell}}

\cmsNoteHeader{HIG-17-011}
\title{Constraints on anomalous Higgs boson couplings using production and decay information in the four-lepton final state}

\date{\today}

\abstract{
A search is performed for anomalous interactions of the recently discovered Higgs boson using matrix element techniques
with the information from its decay to four leptons and from associated Higgs boson production with two quark jets
in either vector boson fusion or associated production with a vector boson. The data were recorded
by the CMS experiment at the LHC at a center-of-mass energy of 13\TeV and correspond
to an integrated luminosity of 38.6\fbinv.
They are combined with the data collected at center-of-mass energies
of 7 and 8\TeV,
corresponding to integrated luminosities of 5.1 and 19.7\fbinv, respectively.
All observations are consistent with the expectations for the standard model Higgs boson.
}

\hypersetup{%
pdfauthor={CMS Collaboration},%
pdftitle={Constraints on anomalous Higgs boson couplings using production and decay information in the four-lepton final state},%
pdfsubject={CMS},%
pdfkeywords={Higgs, exotic spin, anomalous coupling}}

\maketitle

\section{Introduction}

The observation of a boson with a mass of about 125\GeV
by the ATLAS and CMS Collaborations~\cite{Aad:2012tfa,Chatrchyan:2012xdj,Chatrchyan:2013lba}
is consistent with the prediction of the standard model (SM) Higgs ($\PH$)
boson~\cite{StandardModel67_1, Englert:1964et,Higgs:1964ia,Higgs:1964pj,Guralnik:1964eu,StandardModel67_2,StandardModel67_3}.
It has been established that the spin-parity quantum numbers of the $\PH$ boson are
consistent with $J^{PC}=0^{++}$~\cite{Chatrchyan:2012jja,Chatrchyan:2013mxa,Khachatryan:2014kca,Khachatryan:2016tnr,
Khachatryan:2015mma,Aad:2013xqa,Aad:2015mxa,Aad:2016nal}.
However, the data still leave room for anomalous interactions or $C\!P$ violation
in the interactions of the $\PH$ boson.
The kinematics of leptons ($\ell=\mu^\pm$ and $\Pe^\pm$)
from $\PH\to \Z\Z / \Z\gamma^* / \gamma^*\gamma^* \to 4\ell$ decays
(through virtual photons or \Z bosons),
of quark jets produced in association with the $\PH$ boson in
vector boson fusion (VBF), and of the decays of $\Z$ or $\PW$ bosons produced in association with $\PH$ ($\V\PH$)
allow studies of anomalous interactions of the $\PH$
boson~\cite{Nelson:1986ki,Soni:1993jc,Plehn:2001nj,Choi:2002jk,Buszello:2002uu,Godbole:2007cn,
Hagiwara:2009wt,Gao:2010qx,DeRujula:2010ys,Christensen:2010pf,Bolognesi:2012mm,Ellis:2012xd,
Chen:2012jy,Artoisenet:2013puc,Anderson:2013afp,Chen:2013waa,Gonzalez-Alonso:2014eva,Greljo:2015sla}.

The CMS Collaboration analyzed the data collected at the CERN LHC
at center-of-mass energies of 7 and 8\TeV (Run\,1),
corresponding to integrated luminosities of $\usedLumiAnofb$ and $\usedLumiB$,
measuring the spin-parity properties of the $\PH$ boson and searching for anomalous $\PH\V\V$ couplings
using the $\PH$ boson's decay modes to two electroweak gauge bosons~\cite{Khachatryan:2014kca}.
That study focused on testing for the presence of anomalous effects in
$\PH\Z\Z$, $\PH\Z\gamma$, $\PH\gamma\gamma$, and $\PH\PW\PW$ interactions
under spin-zero, -one, and -two hypotheses. The spin-one hypotheses were excluded at greater than 99.999\% confidence level (\CL)
in the $\Z\Z$ and $\PW\PW$ modes; they were also excluded via the Landau--Yang theorem~\cite{Landau,Yang}
by the observation of the $\gamma\gamma$ decay mode with 5.7$\sigma$ significance.
The spin-two boson hypothesis with gravity-like minimal couplings was excluded at 99.87\% \CL,
and nine other possible hypotheses of spin-two tensor structure of $\PH\V\V$ interactions
were excluded at 99\% \CL or higher.
Given the exclusion of the spin-one and -two scenarios,
constraints were set on the contribution of eleven anomalous couplings to the
$\PH\Z\Z$, $\PH\Z\gamma$, $\PH\gamma\gamma$, and $\PH\PW\PW$ interactions under the hypothesis of a spin-zero state.
Among others, these results constrained a $C\!P$-violation parameter $f_{a3}$,
the fractional pseudoscalar cross section in the $\PH\to\Z\Z$ channel,
which will be described in more detail in Section~\ref{sec:Pheno}.
The pure pseudoscalar hypothesis was excluded at 99.98\% \CL, and the limit $f_{a3}<0.43$ was set at 95\% \CL.
Similar results, for a smaller number of parameters and fewer exotic-spin models, were obtained by ATLAS~\cite{Aad:2015mxa}.

All the above studies considered the decay of an on-shell $\PH$ boson to two vector bosons.  The accumulated data
in Run\,1 were not sufficient for precision tests of anomalous interactions in associated production,
in off-shell production, or with fermions. Nonetheless, both CMS~\cite{Khachatryan:2016tnr}
and ATLAS~\cite{Aad:2016nal}
performed analyses of anomalous $\PH\V\V$ interactions in $\V\PH$ and VBF production, respectively.
Finally, the CMS experiment searched for anomalous $\PH\V\V$ interactions in off-shell production
of the $\PH$ boson in $\Pp\Pp\to \PH\to \Z\Z$ with Run\,1 data~\cite{Khachatryan:2015mma}.
Further measurements probing the tensor structure of the $\PH\V\V$ and $\PH\ffbar$ interactions
can test $C\!P$ invariance and, more generally, any small anomalous contributions~\cite{deFlorian:2016spz}.

In this Letter, the analysis approach follows our previous Run\,1 publication~\cite{Khachatryan:2014kca},
expanded in two important ways. Information from the kinematic correlations of quark jets from VBF and $\V\PH$ production
is used together with $\PH\to \Z\Z / \Z\gamma^* / \gamma^*\gamma^* \to 4\ell$ decay information for the first time,
applying the relevant techniques discussed in Ref.~\cite{Anderson:2013afp}.
Moreover, data sets corresponding to integrated luminosities of $\usedLumiCnofb$ and $\usedLumiD$
collected at a center-of-mass energy of 13\TeV in Run\,2 of the LHC
during 2015 and 2016, respectively, are combined with the Run\,1 data,
increasing the data sample of $\PH\to4\ell$ events by approximately a factor of four.

In what follows,
the phenomenology of anomalous $\PH\V\V$ interactions is discussed in Section~\ref{sec:Pheno}.
The CMS detector, reconstruction techniques, and Monte Carlo (MC) simulation are introduced in Section~\ref{sec:CMS}.
Details of the analysis are discussed in Section~\ref{sec:AnalysisStrategyIntro}, and results are presented in Section~\ref{sec:results}.
We summarize in Section~\ref{sec:Summary}.

\section{Phenomenology of anomalous \texorpdfstring{$\PH$}{H} boson interactions}
\label{sec:Pheno}

We assume that the $\PH$ boson couples to two gauge bosons $\V\V$, such as $\Z\Z$, $\Z \gamma$,
$\gamma\gamma$, $\PW\PW$, or $\Pg\Pg$, which in turn couple to quarks or
leptons~\cite{Nelson:1986ki,Soni:1993jc,Plehn:2001nj,Choi:2002jk,Buszello:2002uu,Godbole:2007cn,
Hagiwara:2009wt,Gao:2010qx,DeRujula:2010ys,Christensen:2010pf,Bolognesi:2012mm,Ellis:2012xd,
Chen:2012jy,Artoisenet:2013puc,Anderson:2013afp,Chen:2013waa}.
Three general tensor structures that are allowed by Lorentz symmetry are tested.
Each term includes a form factor $F_i(q_1^2,q_2^2)$,
where $q_1$ and $q_2$ are the four-momenta of the two difermion states, such as $\Pep\Pem$ and $\PGmp\PGmm$ in the
$\PH\to \Pep\Pem\PGmp\PGmm$ decay.
The $\PH$ boson coupling to fermions is assumed not to be mediated by a new heavy state $\V^\prime$,
generating the so-called contact terms~\cite{Gonzalez-Alonso:2014eva,Greljo:2015sla}.
We therefore study the process $\PH\to\V\V\to4\mathrm{f}$ and the equivalent processes in production,
rather than $\PH\to\V\V^\prime\to4\mathrm{f}$ or equivalent processes.
Nonetheless, those contact terms are equivalent to the anomalous $\PH\V\V$ couplings already tested
using the $f_{\Lambda1}$  and $f_{\Lambda1}^{\Z\gamma}$ parameters,
defined below.
It is assumed that all lepton and quark couplings to vector bosons follow the SM predictions.
Relaxing this requirement would be equivalent to allowing the contact terms to vary with flavor,
which would result in too many unconstrained parameters to be tested with the present amount of data.
Only the lowest order operators, or lowest order terms in the $(q_j^2/\Lambda^2)$ form-factor expansion, are tested,
where $\Lambda$ is an energy scale of new physics.

Anomalous interactions of a spin-zero $\PH$ boson with two spin-one gauge bosons $\V\V$, such as $\Z\Z$,
$\Z \gamma$, $\gamma\gamma$, $\PW\PW$, and $\Pg\Pg$, are parameterized with a scattering amplitude that
includes three tensor structures with expansion of coefficients up to $(q^2/\Lambda^2)$:
\ifthenelse{\boolean{cms@external}}{
\begin{multline}
A(\PH\V\V) \sim
\left[ a_{1}^{\V\V}
+ \frac{\kappa_1^{\V\V}q_{1}^2 + \kappa_2^{\V\V} q_{2}^{2}}{\left(\Lambda_{1}^{\V\V} \right)^{2}}
\right]
m_{\V1}^2 \epsilon_{\V1}^* \epsilon_{\V2}^*  \\
+ a_{2}^{\V\V}  f_{\mu \nu}^{*(1)}f^{*(2),\mu\nu}
+ a_{3}^{\V\V}   f^{*(1)}_{\mu \nu} {\tilde f}^{*(2),\mu\nu},
\label{eq:formfact-fullampl-spin0}
\end{multline}
}{
\begin{equation}
A(\PH\V\V) \sim
\left[ a_{1}^{\V\V}
+ \frac{\kappa_1^{\V\V}q_{1}^2 + \kappa_2^{\V\V} q_{2}^{2}}{\left(\Lambda_{1}^{\V\V} \right)^{2}}
\right]
m_{\V1}^2 \epsilon_{\V1}^* \epsilon_{\V2}^*  \\
+ a_{2}^{\V\V}  f_{\mu \nu}^{*(1)}f^{*(2),\mu\nu}
+ a_{3}^{\V\V}   f^{*(1)}_{\mu \nu} {\tilde f}^{*(2),\mu\nu},
\label{eq:formfact-fullampl-spin0}
\end{equation}
}
where $q_{i}$, $\epsilon_{{\V}i}$, and $m_{\V1}$ are the four-momentum, polarization vector, and pole mass
of a gauge boson, $f^{(i){\mu \nu}} = \epsilon_{{\V}i}^{\mu}q_{i}^{\nu} - \epsilon_{{\V}i}^\nu q_{i}^{\mu}$,
${\tilde f}^{(i)}_{\mu \nu} = \frac{1}{2} \epsilon_{\mu\nu\rho\sigma} f^{(i),\rho\sigma}$~\cite{Khachatryan:2014kca,Anderson:2013afp},
and $a_i^{\V\V}$ and $\kappa_i^{\V\V}/\left(\Lambda_1^{\V\V}\right)^2$ are parameters to be determined from data.

In Eq.~(\ref{eq:formfact-fullampl-spin0}), the only leading tree-level contributions are $a_{1}^{\Z\Z}\ne 0$ and $a_{1}^{\PW\PW} \ne 0$,
and we assume custodial symmetry, so that $a_{1}^{\Z\Z}=a_{1}^{\PW\PW}$. The rest of the couplings
are considered anomalous contributions.  Tiny anomalous terms arise in the SM due to loop effects,
and new, beyond standard model (BSM) contributions could make them larger.
The SM values of those couplings are not yet accessible experimentally.
Considerations of gauge invariance and symmetry between two identical bosons require
$\kappa_1^{\Z\Z}=\kappa_2^{\Z\Z}=-\exp({i\phi^{\Z\Z}_{\Lambda{1}}})$,
$\kappa_{1,2}^{\gamma\gamma}=\kappa_{1,2}^{\Pg\Pg}=\kappa_1^{\Z\gamma}=0$,
and $\kappa_2^{\Z\gamma}=-\exp({i\phi^{\Z\gamma}_{\Lambda{1}}})$,
where $\phi^{\V\V}_{\Lambda1}$ is the phase of the corresponding coupling.
The $a^{\Z\gamma}_{2,3}$ and $a^{\gamma\gamma}_{2,3}$ terms were tested in the Run\,1 analysis~\cite{Khachatryan:2014kca},
but have tighter constraints from on-shell photon measurements in $\PH\to \Z\gamma$ and $\gamma\gamma$.
We therefore do not repeat those measurements.
The $\PH\PW\PW$ couplings appear in VBF and $\PW\PH$ production.
We relate those couplings to the $\PH\Z\Z$ measurements assuming $a_{i}^{\PW\PW}=a_{i}^{\Z\Z}$
and drop the $\Z\Z$ labels in what follows.  Four anomalous couplings are left to be tested:
$a_2$, $a_3$, $\kappa_2/\Lambda_1^2$, and $\kappa_2^{\Z\gamma}/\left(\Lambda_1^{\Z\gamma}\right)^2$.
The generic notation $a_i$ refers to all four of these couplings, as well as the SM coupling $a_1$.

\begin{figure*}[tbhp]
\centering
\includegraphics[width=0.32\textwidth]{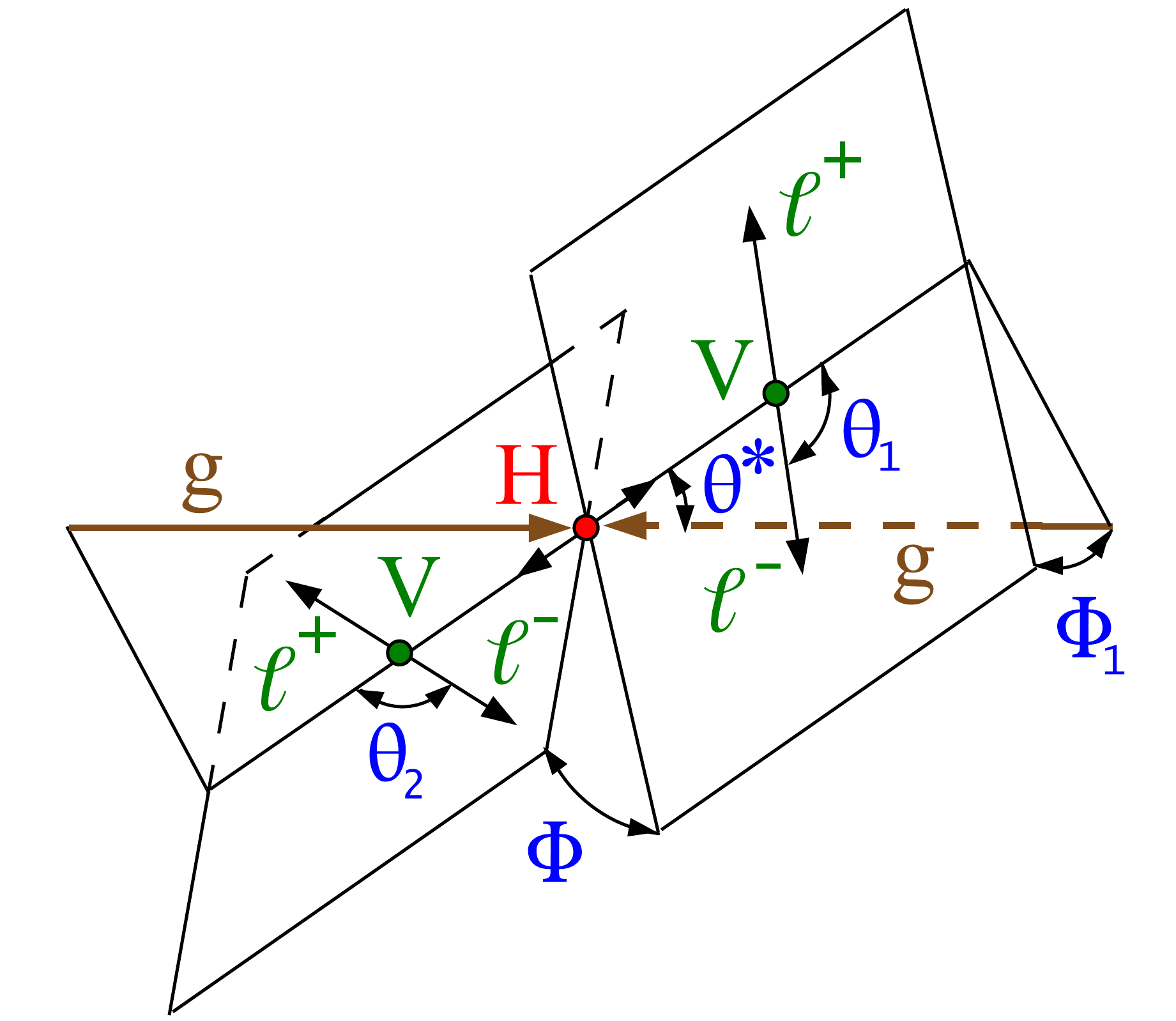}
\includegraphics[width=0.32\textwidth]{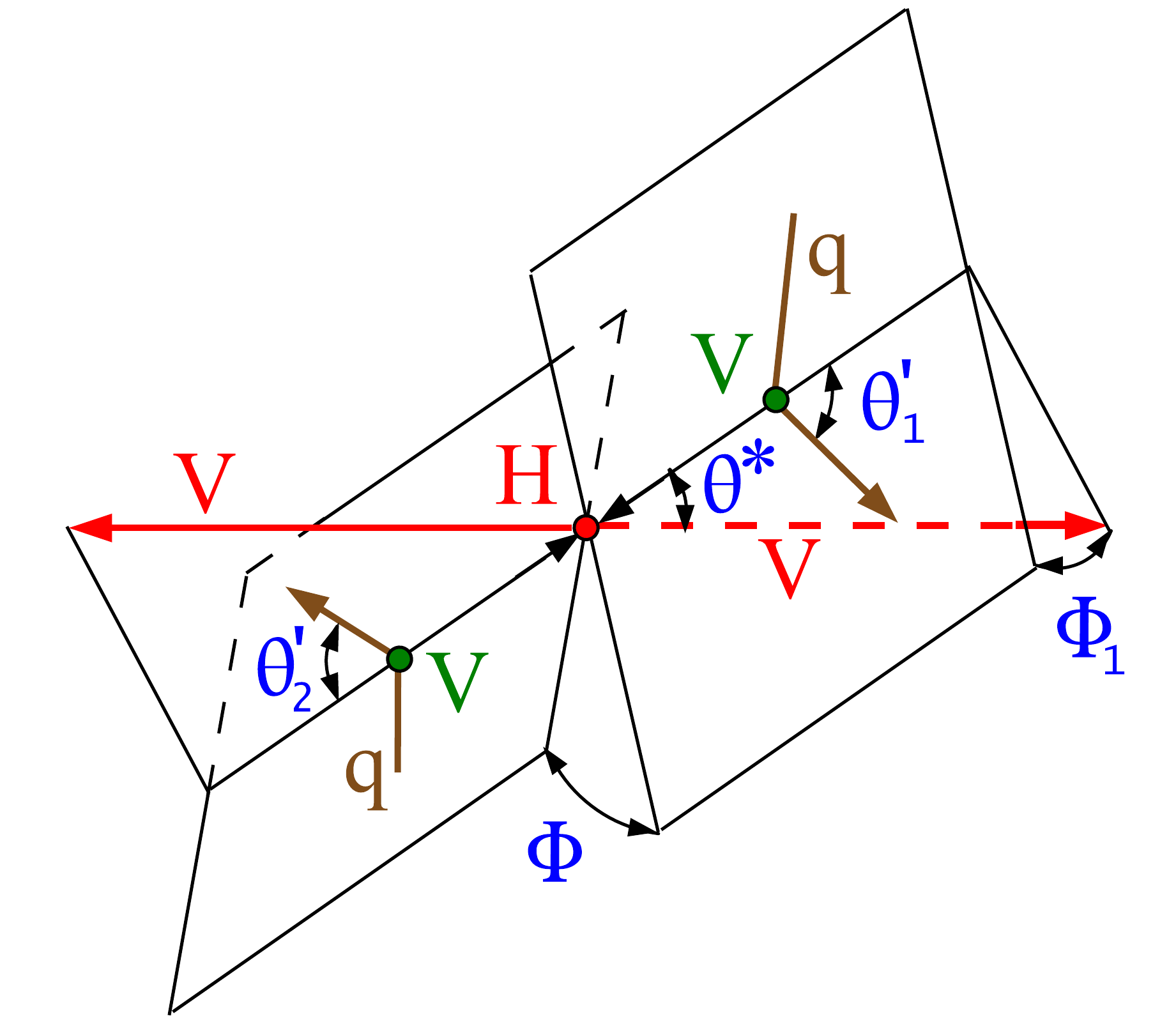}
\includegraphics[width=0.32\textwidth]{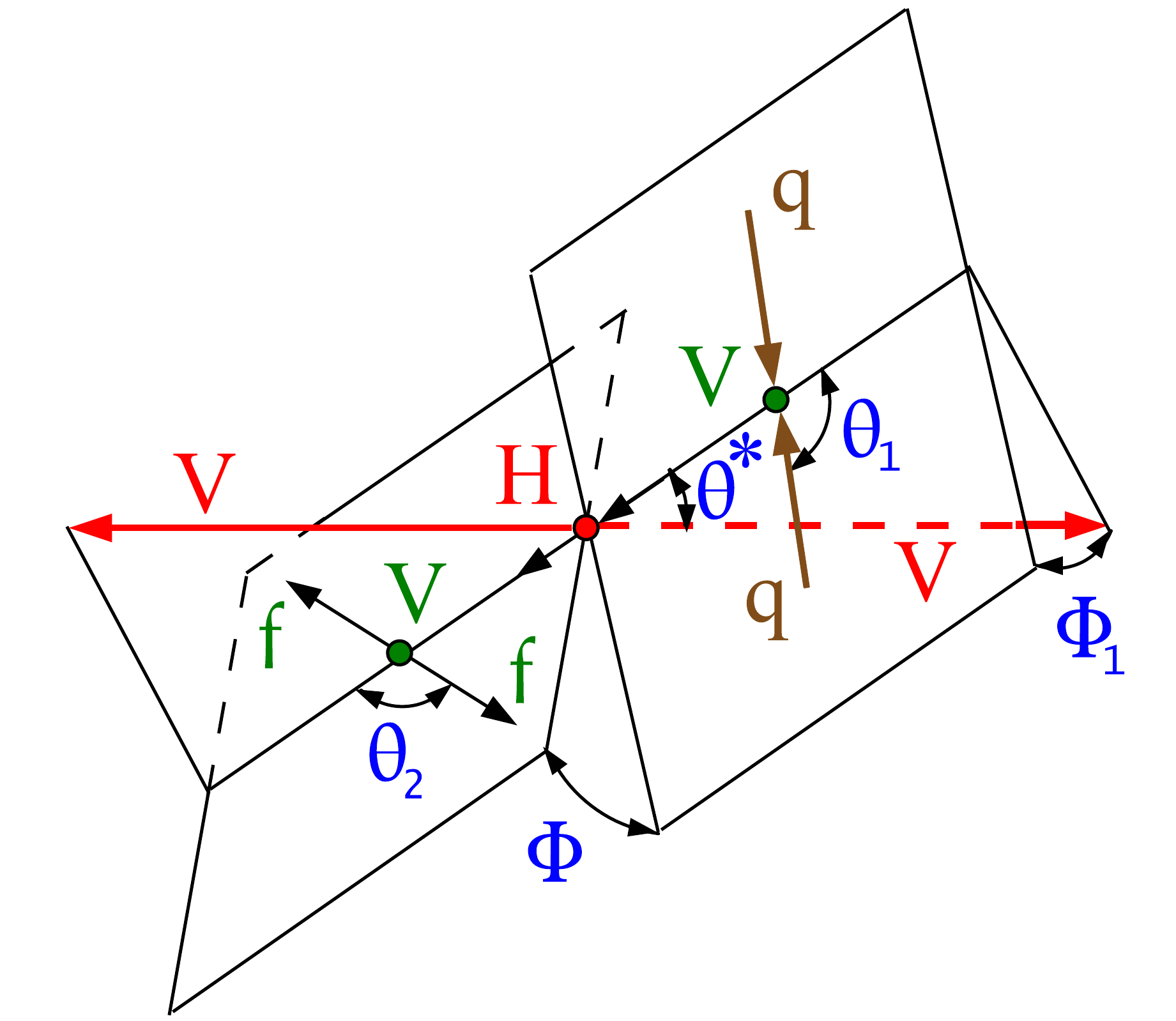}
\caption{
Illustration of $\PH$ boson production and decay in three topologies:
gluon fusion $\Pg\Pg\to \PH \to \V\V \to 4\ell$ (left);
vector boson fusion $\qq\to \V\V (\qq) \to \PH(\qq) \to \V\V(\qq)$ (middle);
and associated production $\qq\to \V \to \V\PH\to  (\ff)\PH \to  (\ff) \V\V$ (right).
In the latter two cases, although the full $\PH$ decay chain is not shown in the figure,
the production and decay $\PH \to \V\V$ may be followed by the same four-lepton decay shown
in the first case. The five angles shown in blue and the invariant masses of the two vector bosons shown in green
fully characterize either the production or the decay chain.
The angles are defined in either the $\PH$ or $\V$ boson rest frames~\cite{Gao:2010qx,Anderson:2013afp}.
}
\label{fig:kinematics}
\end{figure*}

Equation~(\ref{eq:formfact-fullampl-spin0}) parameterizes both the $\PH\to\V\V$ decay and the production of the $\PH$ boson via
either VBF or $\V\PH$.  All three of these processes, which are illustrated in Fig.~\ref{fig:kinematics}, are considered.
While $q_{i}^2$ in the $\PH\to\V\V$ process does not exceed $\left(100\GeV\right)^2$ due to
the kinematic bound, in associated production no such bound exists.
In the present analysis it is assumed that the $q_{i}^2$ range is not restricted within the allowed phase space.

The effective fractional cross sections $f_{ai}$ and phases $\phi_{ai}$ are defined as follows:
\begin{equation}
f_{ai}= {\abs{a_i}^2 \sigma_{i}} \Big/\sum {\abs{a_j}^2 \sigma_{j}}, ~\text{and}~\phi_{ai} = \text{arg}\left({a_{i}}/{a_{1}}\right).
\label{eq:fai}
\end{equation}

This definition of $f_{ai}$ is valid for both the SM coupling $a_1$ and the anomalous couplings,
but there is no need for a separate measurement of $f_{a1}$ because $\sum {f_{ai}}=1$.
The cross sections $\sigma_i$ in Eq.~\ref{eq:fai} are calculated for each corresponding coupling $a_i$.
They are evaluated for the $\PH\to\Z\Z/ \Z\gamma^* /\gamma^*\gamma^*\to2\Pe2\mu$
process, where $a_i=1$ and all other $a_j=0$ in Eq.~(\ref{eq:formfact-fullampl-spin0}).
The resulting ratios are
$\sigma_{1}/\sigma_{3}= 6.53$, $\sigma_{1}/\sigma_{2}= 2.77$,
$\sigma_{1}/{\sigma_{\Lambda1}}= 1.47\times 10^{4}\TeV^{-4}$, and
$\sigma_{1}/{\sigma_{\Lambda1}^{\Z\gamma}}= 5.80\times 10^{3}\TeV^{-4}$.
In the case of the $\PH\Z\gamma$ coupling the requirement $\sqrt{\smash[b]{\abs{q^2_i}}} \ge 4\GeV$ is introduced
in the cross section calculations to avoid infrared divergence.
Equation~(\ref{eq:fai}) can be inverted to recover the coupling ratio,
\begin{equation}
\left|\frac{a_i}{a_1}\right|=\sqrt{\frac{f_{ai}}{f_{a1}}}\,\sqrt{\frac{\sigma_1}{\sigma_i}}.
\label{eq:fai-inverse}
\end{equation}

It is convenient to measure the effective cross-section ratios
($f_{ai}$) rather than the anomalous couplings themselves ($a_i$).
First of all, most systematic uncertainties cancel in the ratio.
Moreover, the effective fractions are conveniently bounded by 0 and 1 and do not depend on the
normalization convention in the definition of the couplings.
Until the effects of interference become important, the statistical uncertainties in these measurements
scale with the integrated luminosity as $1/\sqrt{\mathcal{L}}$, in the same way as cross section measurements.
The $f_{ai}$ values have a simple interpretation as the fractional size of the BSM contribution for
the $\PH\to\Z\Z / \Z\gamma^* /\gamma^*\gamma^* \to2\Pe2\mu$ decay.
For example, $f_{ai}=0$ indicates a pure SM Higgs boson,
$f_{ai}=1$ gives a pure BSM particle, and $f_{ai}=0.5$ means that the two couplings contribute equally
to the $\PH\to\Z\Z/ \Z\gamma^* /\gamma^*\gamma^*\to2\Pe2\mu$ process.
In particular, $f_{a3}$ is the fractional pseudoscalar cross section in the $\PH\to\Z\Z\to2\Pe2\mu$ channel.
A value $0<f_{a3}<1$ would indicate $C\!P$ violation, with a possible mixture of scalar
and pseudoscalar states, while $f_{a3}=1$ would indicate that the $\PH$ boson is a pure pseudoscalar resonance,
which has been excluded at 99.98\% \CL~\cite{Khachatryan:2014kca}.

The above approach allows a general test of the kinematic distributions
associated with the couplings of $\PH$ to 4 fermions, whether in the decay or in the
associated production channels, as shown in Fig.~\ref{fig:kinematics}.
If deviations from the SM are detected,
a more detailed study of the $(q_j^2/\Lambda^2)$ form-factor expansion can be performed, eventually providing
a measurement of the double-differential cross section for each tested tensor structure.
Under the assumption that the couplings are constant and real (\ie, $\phi_{ai}=0$ or~$\pi$),
the above formulation is equivalent to an effective Lagrangian~\cite{Khachatryan:2014kca}.
It is also equivalent to the formulation involving contact terms~\cite{Gonzalez-Alonso:2014eva,Greljo:2015sla}
if the contact terms are assumed to satisfy lepton universality.

\section{The CMS detector, simulation, and reconstruction}
\label{sec:CMS}

The $\PH\to 4\ell$ decays are reconstructed in the CMS detector,
which is composed of a silicon pixel and strip tracker, a lead tungstate crystal electromagnetic
calorimeter (ECAL), and a brass and scintillator hadron calorimeter, each composed of a barrel and two endcap sections,
all within a superconducting solenoid of 6\unit{m} internal diameter, providing a magnetic field of 3.8\unit{T}.
Outside the solenoid are the gas-ionization detectors for muon measurements, which are embedded
in the steel flux-return yoke. Extensive forward calorimetry
complements the coverage provided by the barrel and endcap detectors.
Events of interest are selected using a two-tiered trigger system.
A detailed description of the CMS detector,
together with a definition of the coordinate system used and the relevant kinematic variables,
can be found in Ref.~\cite{Chatrchyan:2008zzk}.

A dedicated MC program, \textsc{JHUGen}~7.0.2~\cite{Gao:2010qx,Bolognesi:2012mm,Anderson:2013afp,Gritsan:2016hjl},
is used to simulate the effect of anomalous couplings in the production and decay $\PH\to \Z\Z$ / $\Z\gamma^*$ / $\gamma^*\gamma^*\to4\ell$.
The gluon fusion production of an $\PH$ boson is simulated with the \POWHEG~2.0~\cite{Frixione:2007vw,Bagnaschi:2011tu,Nason:2009ai}
event generator at next-to-leading order (NLO) in quantum chromodynamics (QCD).
The associated gluon fusion production of an $\PH$ boson with two jets is affected by anomalous $\PH\Pg\Pg$ interactions.
These effects are modeled with \textsc{JHUGen}.
It is also found that the NLO QCD effects that are relevant for the analysis of a spin-zero state
are well described by a combination of leading-order (LO) matrix elements and parton showering~\cite{Anderson:2013afp}.
For the SM case, \textsc{JHUGen} simulations at LO in QCD and \POWHEG simulations at NLO in QCD,
with parton showering applied in both cases, are explicitly compared, and no significant differences are found.
Therefore, \textsc{JHUGen} at LO in QCD is adopted for the simulation
of VBF, $\V\PH$, and $\ttbar\PH$ production with anomalous couplings.
The \textsc{MELA} package~\cite{Chatrchyan:2012xdj,Gao:2010qx,Bolognesi:2012mm,Anderson:2013afp,Gritsan:2016hjl}
contains a library of matrix elements from
\textsc{JHUGen} for the $\PH$ boson signal and \MCFM~7.0~\cite{MCFM,Campbell:2011bn,Campbell:2013una} for the SM background
and is used to apply weights to events in any MC sample to model any other set of couplings.

The main background in this analysis, $\qqbar\to\Z\Z/\Z\gamma^*\to 4\ell$, is estimated from simulation with \POWHEG,
with the next-to-NLO (NNLO) K-factor, which is approximately 1.1 at $\mllll=125\GeV$~\cite{CMS-HIG-16-041}, applied to the NLO cross section.
The $\Pg\Pg\to\Z\Z/\Z\gamma^*\to 4\ell$ background process is simulated with \MCFM~7.0, where the Higgs
boson production K-factor at NNLO in QCD, which is approximately 2.3 at $\mllll=125\GeV$,
is applied to the LO cross section~\cite{CMS-HIG-14-002}.
The VBF and triple-gauge-boson ($\V\V\V$) backgrounds are estimated at LO with~\textsc{phantom}~1.2.8~\cite{Ballestrero:2007}.
The parton distribution functions (PDFs) used for all of these samples are NNPDF3.0~\cite{Ball:2011uy}.
All MC samples are interfaced with \PYTHIA~8.212~\cite{Sjostrand2015159} tune CUETP8M1~\cite{Khachatryan:2015pea} for parton
showering and further processed through a dedicated simulation of the CMS detector based on \GEANTfour~\cite{Agostinelli2003250}.

The selection of the $\PH\to 4\ell$ events and associated particles closely follows the methods used in the analyses
of Run\,1~\cite{Khachatryan:2014kca,Chatrchyan:2013mxa} and Run\,2~\cite{CMS-HIG-16-041} data.
The main triggers for this analysis select a pair of leptons passing loose identification and isolation requirements,
with $\pt$ of the leading and subleading electron (muon) at least 23 (17) and 12 (8)\GeV, respectively.
To maximize the signal acceptance, triggers requiring three leptons with lower $\pt$ thresholds and no isolation
requirement are also used, as are isolated single-electron and single-muon triggers with higher $\pt$ thresholds.
Electrons (muons) are reconstructed within the geometrical acceptance defined by $\abs{\eta} < 2.5~(2.4)$
for transverse momentum $\pt > 7\,(5)\GeV$ with an algorithm that combines information from the ECAL
(muon system) and the tracker.
It is required that the ratio of each lepton track's impact parameter in three dimensions,
computed with respect to the chosen primary vertex position, to its uncertainty be less than 4.
The primary vertex is defined as the vertex with the highest sum of $\pt^2$ of physics objects defined by a jet-finding algorithm.
To discriminate prompt leptons from $\cPZ/\gamma^*$ boson decays from those arising from hadron decays within jets,
an isolation requirement for leptons is imposed.
An algorithm is used to collect the final-state radiation (FSR) of leptons.
An FSR photon is associated to the closest selected lepton in the event if its angular separation
from the lepton is below the required threshold, as discussed in Ref.~\cite{CMS-HIG-16-041}.
Three mutually exclusive channels are considered: $\PH\to4\Pe$, $4\Pgm$, and $2\Pe 2\Pgm$.
At least two leptons are required to have $\pt > 10\GeV$, and at least one is required to have $\pt > 20\GeV$.
All four pairs of oppositely charged leptons that can be built with the four leptons, irrespective of flavor, are required
to satisfy $m_{\ell^{+}\ell'^{-}} > 4\GeV$. The $\cPZ/\gamma^*$ candidates are required to satisfy the condition $12\GeV < \mlplm < 120\GeV$;
the invariant mass of at least one of the $\cPZ/\gamma^*$ candidates must be larger than 40\GeV.
The four-lepton invariant mass $\mllll$ must be between 105 and 140\GeV.

Jets are reconstructed using the particle-flow (PF) algorithm \cite{CMS-PRF-14-001},
with PF candidates clustered by the anti-\kt
algorithm~\cite{Cacciari:2008gp, Cacciari:2011ma} with a distance parameter of 0.4, and with the constraint
that the charged particles be compatible with the primary vertex. The jet momentum is determined as the vectorial sum of all
PF candidate momenta in the jet. Jets must satisfy $\pt>30\GeV$ and $\abs{\eta}<4.7$ and be separated from all selected lepton
candidates and any selected FSR photons by an angular distance $\Delta R(\ell/\cPgg,\text{jet})>0.4$,
where the angular distance between two particles
$i$ and $j$ is $\Delta R(i,j) = \sqrt{\smash[b]{(\eta^i-\eta^j)^{2} + (\phi^i-\phi^j)^{2}}}$.

\section{Analysis techniques}
\label{sec:AnalysisStrategyIntro}

The full kinematic information from each event is extracted using the matrix element calculations in the \textsc{mela} package.
For either the $\PH$ boson decay or associated production with two jets, up to seven kinematic observables,
five angles and two invariant masses, are defined,
as shown in Fig.~\ref{fig:kinematics}~\cite{Gao:2010qx,Anderson:2013afp}.
In the $2\to 6$ process of associated $\PH$ boson production via either VBF, $\Z\PH$, or $\PW\PH$ and
its subsequent decay to a four-fermion final state, up to 13 independent observables $\vec\Omega$ remain.
In the following, we use either the production kinematics, the decay kinematics, or both, as appropriate.
The $\ptvec$ of the system of the $\PH$ boson and two jets, which would appear at NLO in QCD,
is not included in the input observables in order to reduce associated QCD uncertainties.
The \textsc{MELA} approach retains all relevant kinematic information in a minimal set of discriminants
$\mathcal{D}$, computed from ratios of probabilities $\mathcal{P}$.  We use two types of discriminants,
\begin{equation}
\mathcal{D}_\text{alt}\left(\vec\Omega\right) = \frac{\mathcal{P}_\text{sig}\left(\vec\Omega\right) }{\mathcal{P}_\text{sig}\left(\vec\Omega\right) +\mathcal{P}_\text{alt}\left(\vec\Omega\right) }
\label{eq:melaD}
\end{equation}
and
\begin{equation}
\mathcal{D}_\text{int}\left(\vec\Omega\right) = \frac{\mathcal{P}_\text{int}\left(\vec\Omega\right) }{\mathcal{P}_\text{sig}\left(\vec\Omega\right) +\mathcal{P}_\text{alt}\left(\vec\Omega\right) },
\label{eq:melaDint}
\end{equation}
where ``sig'' stands for the SM signal; ``alt'' denotes an alternative hypothesis~\cite{Bolognesi:2012mm},
which could be background (``bkg''), an alternative $\PH$ boson production mechanism (``2jet''), or an alternative $\PH$ boson coupling model (``$a_i$'');
and ``int'' represents the contribution to the probability from the interference between ``sig'' and ``alt''~\cite{Anderson:2013afp}.
By the Neyman-Pearson lemma~\cite{Neyman289}, the $\mathcal{D}_\text{alt}$ discriminant contains all the information
available from the kinematics to separate the SM signal hypothesis from the
alternative hypothesis.  Because all intermediate hypotheses are a linear combination of the SM hypothesis
and the alternative hypothesis,
the combination of $\mathcal{D}_\text{alt}$ with $\mathcal{D}_\text{int}$
also contains all the information available to separate the interference component.  The discriminants used
in this analysis are summarized in Table~\ref{table:categories} and described in more detail below.

\begin{table*}[tbh]
\begin{center}
\topcaption{
Summary of the three production categories in the analysis of 2016 data.
The discriminants $\mathcal{D}$ are calculated from Eqs.~(\ref{eq:melaD}) and (\ref{eq:melaDint}),
as discussed in more detail in the text.
For each analysis, the appropriate BSM model is considered in the definition of the categories:
$f_{a3}=1$, $f_{a2}=1$, $f_{\Lambda1}=1$, or $f^{\Z\gamma}_{\Lambda1}=1$.
Three observables (abbreviated as obs.) are listed for each analysis and for each category.
They are described in more detail later in the text.
 }
 \renewcommand*{\arraystretch}{1.8}
\begin{tabular}{lccc}
\hline
   Category              & VBF-jet & $\V\PH$-jet  & Untagged \\
\hline
  Target         &    $\PQq{\PQq}^\prime \V\V\to \PQq{\PQq}^\prime \PH\to (\cmsSymbolFace{j}\cmsSymbolFace{j})(4\ell)$            &      $\qqbar\to \V\PH\to (\cmsSymbolFace{j}\cmsSymbolFace{j})(4\ell)$     &  $\PH\to 4\ell$   \\
  Selection
              &     $ \mathcal{D}_\text{2jet}^\text{VBF}$ or $ \mathcal{D}_\text{2jet}^{\text{VBF}, \text{BSM}} >0.5$
             &     $ \mathcal{D}_\text{2jet}^{\Z\PH}$  or $ \mathcal{D}_\text{2jet}^{\Z\PH, \text{BSM}}$ or
              & not VBF-jet \\[-0.6ex]
              &
             &     $ \mathcal{D}_\text{2jet}^{\PW\PH}$  or $ \mathcal{D}_\text{2jet}^{\PW\PH, \text{BSM}} >0.5$
              & not $\V\PH$-jet \\[-0.6ex]
  \hline
   $f_{a3}$ obs.
               &  $\mathcal{D}_\text{bkg}$, $\mathcal{D}_{0-}^{\text{VBF}\!+\!\text{dec}}$, $\mathcal{D}_{C\!P}^\text{VBF}$
             &  $\mathcal{D}_\text{bkg}$, $\mathcal{D}_{0-}^{\V\PH\!+\!\text{dec}}$, $\mathcal{D}_{C\!P}^{\V\PH}$
             &  $\mathcal{D}_\text{bkg}$, $\mathcal{D}_{0-}^\text{dec}$, $\mathcal{D}_{C\!P}^\text{dec}$  \\
  $f_{a2}$ obs.
               &  $\mathcal{D}_\text{bkg}$, $\mathcal{D}_{0h+}^{\text{VBF}\!+\!\text{dec}}$, $\mathcal{D}_\text{int}^\text{VBF}$
             &  $\mathcal{D}_\text{bkg}$, $\mathcal{D}_{0h+}^{\V\PH\!+\!\text{dec}}$, $\mathcal{D}_\text{int}^{\V\PH}$
             &  $\mathcal{D}_\text{bkg}$, $\mathcal{D}_{0h+}^\text{dec}$, $\mathcal{D}_\text{int}^\text{dec}$  \\
  $f_{\Lambda1}$ obs.
               &  $\mathcal{D}_\text{bkg}$, $\mathcal{D}_{\Lambda1}^{\text{VBF}\!+\!\text{dec}}$, $\mathcal{D}_{0h+}^{\text{VBF}\!+\!\text{dec}}$
             &  $\mathcal{D}_\text{bkg}$, $\mathcal{D}_{\Lambda1}^{\V\PH\!+\!\text{dec}}$, $\mathcal{D}_{0h+}^{\V\PH\!+\!\text{dec}}$
             &  $\mathcal{D}_\text{bkg}$, $\mathcal{D}_{\Lambda1}^\text{dec}$,  $\mathcal{D}_{0h+}^\text{dec}$  \\
   $f^{\Z\gamma}_{\Lambda1}$ obs.
               &  $\mathcal{D}_\text{bkg}$,  $\mathcal{D}^{\Z\gamma,{\text{VBF}\!+\!\text{dec}}}_{\Lambda1}$,  $\mathcal{D}_{0h+}^{\text{VBF}\!+\!\text{dec}}$
             &  $\mathcal{D}_\text{bkg}$,  $\mathcal{D}^{\Z\gamma,{\V\PH\!+\!\text{dec}}}_{\Lambda1}$,  $\mathcal{D}_{0h+}^{\V\PH\!+\!\text{dec}}$
             &  $\mathcal{D}_\text{bkg}$, $\mathcal{D}^{\Z\gamma,\text{dec}}_{\Lambda1}$,  $\mathcal{D}_{0h+}^\text{dec}$  \\
\hline
\end{tabular}
\label{table:categories}
\end{center}
\end{table*}

The selected events in the 2016 data sample are split into three categories: VBF-jet, VH-jet, and untagged.
The VBF-jet category requires exactly four leptons with either two or three jets of which at most one is $\cPqb$ quark flavor-tagged,
or at least four jets and no $\cPqb$-tagged jets. The $\V\PH$-jet category requires exactly four leptons and two or more jets;
if there are four or more jets, none of them should be $\cPqb$ tagged.
The requirements on the number of $\cPqb$-tagged jets are applied to reduce cross-feed from $\ttbar\PH$ production.
In order to separate the target production mode for each category from gluon fusion production,
the requirement $\mathcal{D}_\text{2jet}>0.5$ is applied following Eq.~(\ref{eq:melaD}),
where $\mathcal{P}_\text{sig}$ corresponds to the signal probability for the VBF ($\Z\PH$ or $\PW\PH$)
production hypothesis in the  VBF-jet ($\V\PH$-jet) category, and $\mathcal{P}_\text{alt}$ corresponds to the gluon
fusion production of the $\PH$ boson in association with two jets.
In each of the four $f_{ai}$ analyses,
the requirement $\mathcal{D}_\text{2jet}>0.5$ is tested with both the $f_{ai}=0$ and $f_{ai}=1$
signal hypotheses in $\mathcal{P}_\text{sig}$.  Thus,
this categorization differs slightly in the four analyses.
The two highest $\pt$ jets are used in the
calculation of the matrix elements. All events not assigned to the VBF-jet
or $\V\PH$-jet categories are assigned to the untagged category.
The above requirements are summarized in Table~\ref{table:categories}.
Due to the small size of the 2015 data sample, those events were not categorized
and were all treated as untagged, as was done in the analysis of 2011 and 2012 data~\cite{Khachatryan:2014kca}.
The expected and observed numbers of events are listed in Table~\ref{table:category-yields-fa3}.

\begin{table*}[bth]
\begin{center}
\topcaption{
The numbers of events expected for the SM (or $f_{a3}=1$, in parentheses) for different signal and background modes
and the total observed numbers of events across the three $f_{a3}$ categories in 2016 and 2015 data.
}
\begin{tabular}{lcccc}
\hline
 & VBF-jets & $\V\PH$-jets & Untagged & ~~2015~~\\\hline
VBF signal & 2.4 (1.6) & 0.1 (0.1) & 2.2 (0.3) & 0.4 (0.2)\\
$\Z\PH$ signal & 0.1 (0.2) & 0.3 (0.5) & 0.7 (1.0) & 0.1 (0.1)\\
$\PW\PH$ signal & 0.1 (0.3) & 0.3 (1.0) & 0.8 (2.2) & 0.1 (0.3)\\
$\Pg\Pg\to\PH$ signal & 3.2 (3.3) & 1.9 (2.0) & 49.6 (49.4) & 4.6 (4.6)\\
$\ttbar\PH$ signal & 0.1 (0.1) & 0.0 (0.0) & 0.5 (0.6) & 0.1 (0.1)\\\hline
$\qqbar\to4\ell$ bkg & 0.9 & 1.1 & 56.3 & 5.4\\
$\Pg\Pg\to4\ell$ bkg & 0.1 & 0.1 & 5.5 & 0.5\\
VBF/$\V\V\V$ bkg & 0.1 & 0.0 & 0.4 & 0.0\\
$\Z\!+\!\X$ bkg & 3.6 & 2.0 & 29.1 & 1.7\\\hline
Total expected & 10.7 & 5.8 & 145.2 & 12.9\\
Total observed & 11 & 2 & 145 & 11
\\\hline
\end{tabular}
\label{table:category-yields-fa3}
\end{center}
\end{table*}

We perform an unbinned extended maximum likelihood fit to the events split into the categories according
to the lepton flavor and production topology.
An independent fit is performed for each parameter defined in Table~\ref{tab:summary_spin0}.
In each category of events, three observables $\vec{\mathcal{D}}=\{ \mathcal{D}_\text{bkg}, \mathcal{D}_{ai}, \mathcal{D}_\text{int} \}$
are defined following Eqs.~(\ref{eq:melaD}) and (\ref{eq:melaDint}), as summarized in Table~\ref{table:categories}.

\begin{figure*}[tbh]
\centering
\includegraphics[width=1.\textwidth]{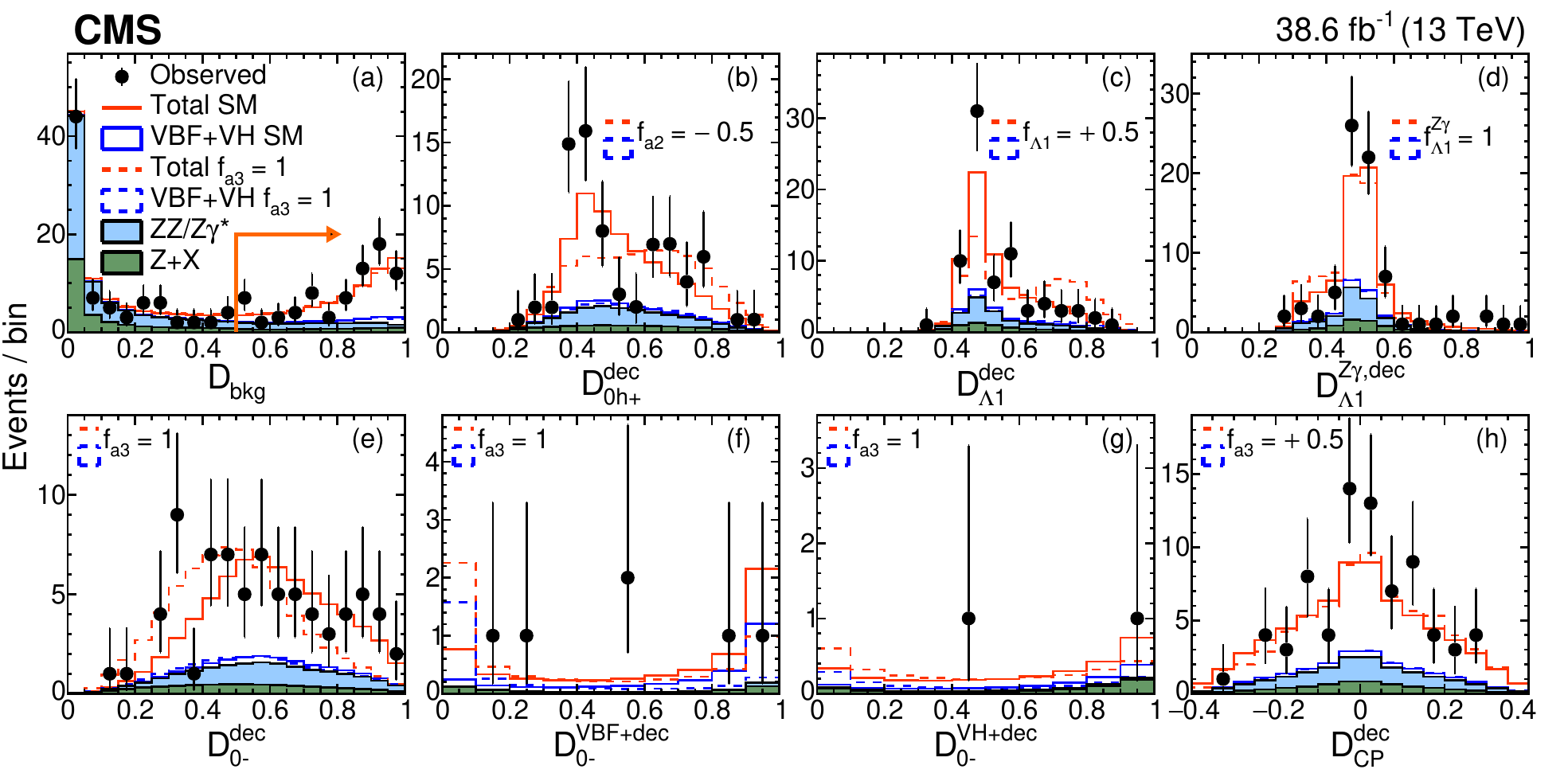}
\caption{
Distributions of $\mathcal{D}_\text{bkg}$ (a) for all events in Run\,2; $\mathcal{D}_{0h+}$ (b), $\mathcal{D}_{\Lambda1}$ (c) ,
$\mathcal{D}_{\Lambda1}^{\Z\gamma}$ (d), $\mathcal{D}_{0-}$ (e), and $\mathcal{D}_{C\!P}$ (h) for the untagged and 2015 events;
$\mathcal{D}_{0-}$ in the VBF-jet (f) and $\V\PH$-jet (g) categories.
The arrow in (a) indicates the requirement $\mathcal{D}_\text{bkg}>0.5$, used to suppress background on all other plots.
Points with error bars show data and histograms show expectations for background and signal, as indicated in the legend in (a).
The dashed lines show expectations for BSM hypotheses, as indicated in the individual legends.
\label{fig:discriminants}
}
\end{figure*}

The first observable, $\mathcal{D}_\text{bkg}$ (shown in Fig.~\ref{fig:discriminants}~(a)),
is common to all events and is designed to separate the signal from the dominant $\qqbar\to4\ell$ background,
for which $\mathcal{P}_\text{bkg}$ is calculated. The signal and background probabilities
include both the matrix element probability based on lepton kinematics and
the $m_{4\ell}$ probability parameterization extracted from simulation of detector effects.
The signal $m_{4\ell}$ parameterization assumes that $m_\PH=125$\GeV.

The second observable, $\mathcal{D}_{ai}$,
separates the SM hypothesis $f_{ai}=0$ from the alternative hypothesis $f_{ai}=1$.
It is defined following Eq.~(\ref{eq:melaD}), with $\mathcal{P}_\text{sig}$ calculated for
$f_{ai}=0$ and $\mathcal{P}_\text{alt}$ for the alternative $\PH$ boson coupling hypothesis with $f_{ai}=1$.
In the untagged category the probabilities are calculated using only the decay information,
but in the VBF-jet and $\V\PH$-jet categories both the production and decay probabilities
are used, with the matrix elements calculated for either $\text{VBF}\times\text{decay}$ or $(\Z\PH+\PW\PH)\times\text{decay}$, respectively.
This observable is called $\mathcal{D}_\mathrm{0-}$ in the $f_{a3}$,
$\mathcal{D}_\mathrm{0h+}$ in the $f_{a2}$, $\mathcal{D}_{\Lambda1}$ in the $f_{\Lambda1}$, and
$\mathcal{D}_{\Lambda1}^{\Z\gamma}$ in the $f_{\Lambda1}^{\Z\gamma}$ analyses~\cite{Khachatryan:2014kca}.
Superscripts are added to the discriminant name to indicate the processes used to calculate the matrix elements:
either dec, $\text{VBF}\!+\!\text{dec}$, or $\V\PH\!+\!\text{dec}$ to denote decay, $\text{VBF}\times\text{decay}$,
or $(\Z\PH+\PW\PH)\times\text{decay}$, respectively.
Distributions of $\mathcal{D}_\mathrm{0-}$
in the three categories are shown in Fig.~\ref{fig:discriminants} (e), (f), (g).
Figure~\ref{fig:discriminants} (b), (c), (d) also shows the distributions of $\mathcal{D}_\mathrm{0h+}$,
$\mathcal{D}_{\Lambda1}$, and $\mathcal{D}_{\Lambda1}^{\Z\gamma}$, respectively, for the untagged events.

The third observable, $\mathcal{D}_\text{int}$
from Eq.~(\ref{eq:melaDint}), separates the interference
of the two amplitudes corresponding to the SM coupling and the alternative $\PH$ boson coupling model.
In the case of the $f_{a3}$ analysis, this observable is called $\mathcal{D}_{C\!P}$ because if $C\!P$ is violated
it would exhibit a distinctive forward-backward asymmetry between $\mathcal{D}_{C\!P}>0$ and $\mathcal{D}_{C\!P}<0$,
as illustrated in Fig.~\ref{fig:discriminants} (h)
for the untagged category of events.
In the untagged category, decay information is used in the calculation of $\mathcal{D}_\text{int}$.
In the VBF-jet and $\V\PH$-jet categories, production information is used.
As in the case of $\mathcal{D}_{ai}$, superscripts indicate which processes were used to calculate the matrix elements.
In the $f_{\Lambda1}$ and $f_{\Lambda1}^{\Z\gamma}$ analyses,
the interference discriminant does not provide additional separation,
and $\mathcal{D}_\mathrm{0h+}$ is used as the third observable.

In the likelihood fit, the signal probability density function (pdf) is parameterized for each production mode
and in each category as
\begin{eqnarray}
P_\text{sig}\left(\vec{\mathcal{D}}; f_{ai},\phi_{ai} \right) \propto
\sum_n \left|\frac{a_i}{a_1}\right|^n  \mathcal{T}_{n}\left(\vec{\mathcal{D}}\right) \cos^n(\phi_{ai})\,,
\label{eq:probability}
\end{eqnarray}
where $\mathcal{T}_{n}$ is the three-dimensional template probability obtained from MC simulation,
$|{a_i}/{a_1}|$ is calculated from $f_{ai}$ through Eq.~(\ref{eq:fai-inverse}),
and $\cos(\phi_{ai})=\pm1$.
The sum runs over five values $n=0,\ldots,4$ in the case of VBF and $\V\PH$,
where the $\PH\V\V$ coupling appears on both the production and decay sides,
and over three contributions $n=0,1,$ and 2 for the other signal modes.
The background pdf is also parameterized with templates extracted from simulation, except for the
reducible background, $\Z\!+\!\X$, which is dominated by the $\Z\!+\!\text{jets}$ process but also includes the
$\ttbar\!+\!\text{jets}$, $\cPZ\gamma\!+\!\text{jets}$, $\PW\cPZ\!+\!\text{jets}$, and $\PW\PW\!+\!\text{jets}$ processes.
The $\Z\!+\!\X$ background is estimated using
independent control regions in data with loose identification requirements on two leptons.

The yields of signal events in 2016 data are expressed with two unconstrained parameters $\mu_V$ and $\mu_F$,
which are the ratios of the observed yields to the expectation
in the SM for the production mechanisms driven by the
$\PH\V\V$ couplings (VBF and $\V\PH$) and for the other modes
(gluon fusion and $\ttbar\PH$), respectively.
The signal yield in 2015 data is expressed with a single parameter $\mu_{13\TeV}$, which is
a linear combination of $\mu_V$ and $\mu_F$.
The fit is also performed simultaneously with the 2011 and 2012 data from Ref.~\cite{Khachatryan:2014kca},
where the two signal strength parameters $\mu_{7\TeV}$ and $\mu_{8\TeV}$
are also linear combinations of $\mu_V$ and $\mu_F$ including the effects of the cross section scaling for each value of $f_{ai}$.

Most uncertainties in the signal yields cancel in this analysis because measurements of
anomalous couplings are expressed as relative cross sections.
Statistical uncertainties dominate over any systematic uncertainties in this analysis.
In the decay-only observables the main effects come from lepton momentum uncertainties
and are propagated into the template uncertainties as in the previous analyses~\cite{Khachatryan:2014kca},
where the main effect is on the $m_{4\ell}$ resolution affecting the $\mathcal{D}_\text{bkg}$ parameterization.

The primary new feature in this analysis, compared to Run\,1~\cite{Khachatryan:2014kca},
is the categorization based on jets and the kinematic discriminants using jet information.
Both the shapes and the yields are varied according to uncertainties
obtained from the jet energy variations. In addition, uncertainties in renormalization and factorization scales, PDFs, and the modeling of
hadronization and the underlying event in MC simulation are propagated to the template and relative yield uncertainties.
As part of these studies, comparisons were made between QCD production at NLO and LO, with matched \PYTHIA hadronization
in each case, for the VBF, $\V\PH$, and $\ttbar\PH$ processes.  In all cases, only small differences
were observed. The uncertainties in the migration of signal and background events between categories
amount to 3--13\% for the signal and 4--25\% for the background, depending on the category.
Among the signal processes, the largest uncertainties arise from the prediction of the $\Pg\Pg\to \PH$ yield in the VBF-jet category.
In $\ttbar\PH$ and gluon fusion production, anomalous couplings on the production side are not generally
related to the $\PH\V\V$ anomalous couplings considered here.
There is a negligible effect on the observed distributions with large variations in the couplings.

Backgrounds from the $\qqbar\to 4\ell$, $\Pg\Pg\to 4\ell$, VBF, and $\V\!+\!(4\ell)$ processes are estimated
using MC simulation. Theoretical uncertainties in the background estimation include uncertainties from
the renormalization and factorization scales, the PDFs, and the K-factors described above.
An additional 10\% uncertainty is assigned to the $\Pg\Pg\to 4\ell$
background K-factor to cover potential differences between signal and background.

\section{Results and discussion}
\label{sec:results}

\begin{figure*}[tbh]
\centering
	\includegraphics[width=.4\textwidth]{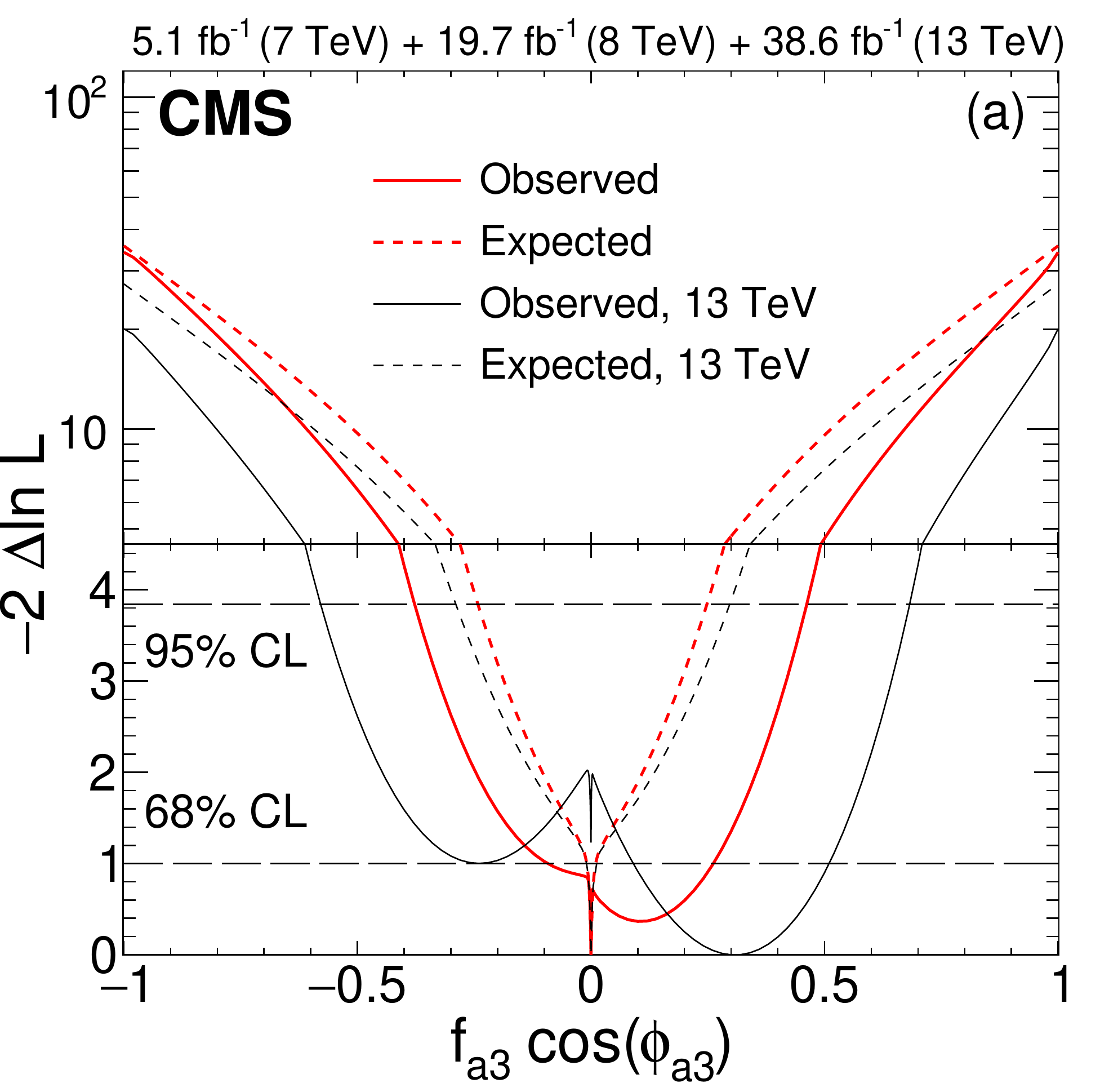}
	\includegraphics[width=.4\textwidth]{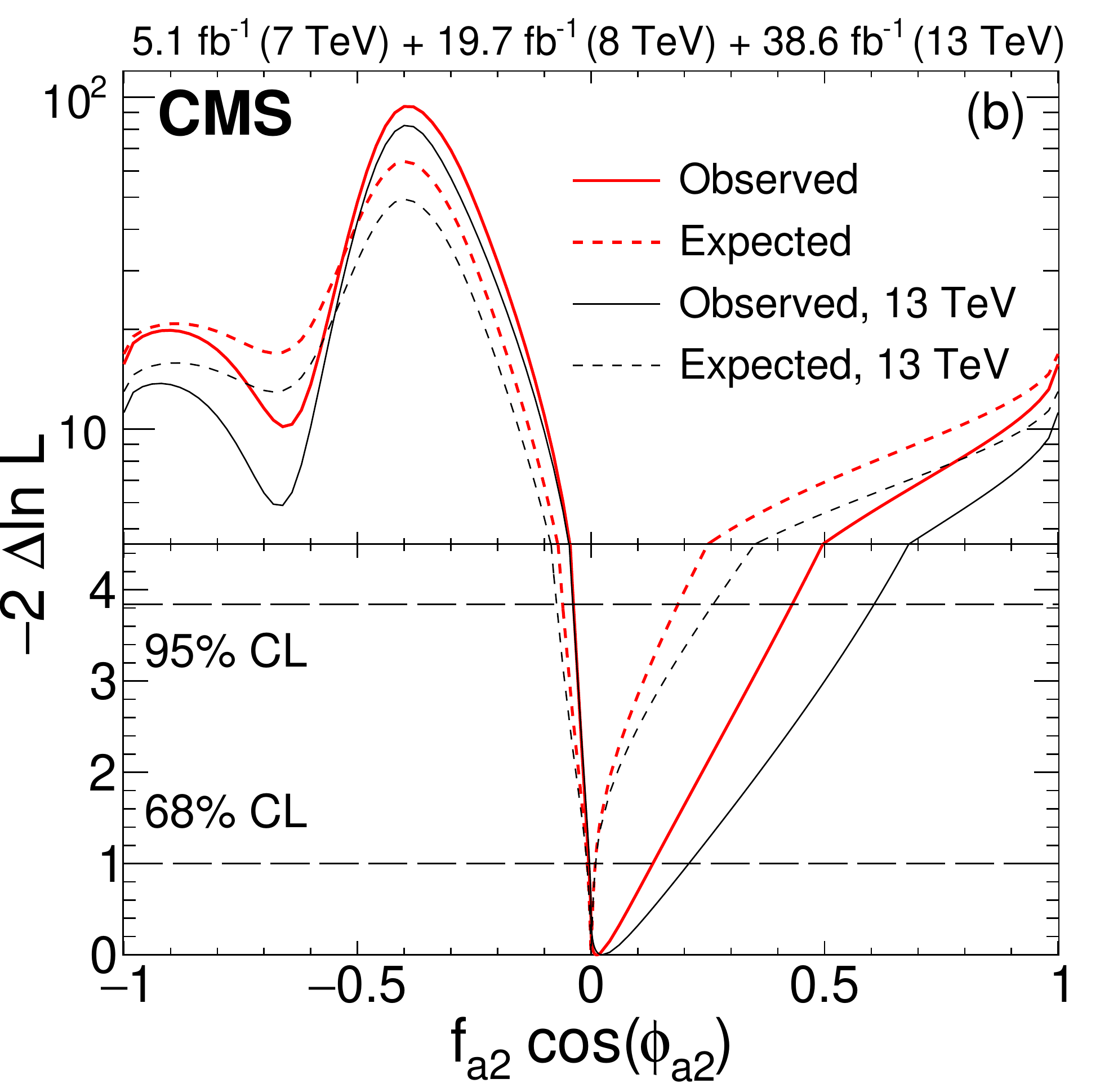}
	\includegraphics[width=.4\textwidth]{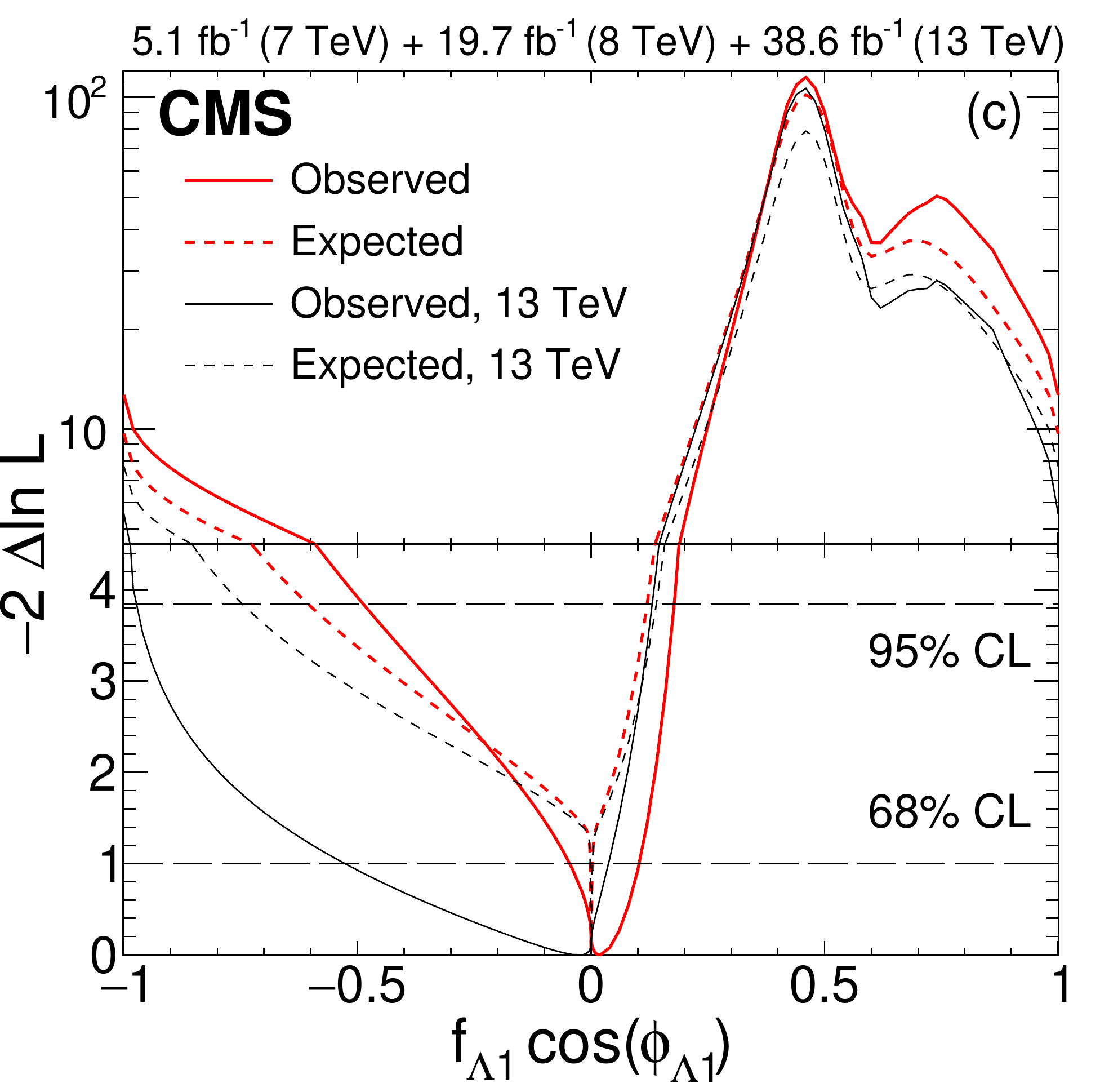}
	\includegraphics[width=.4\textwidth]{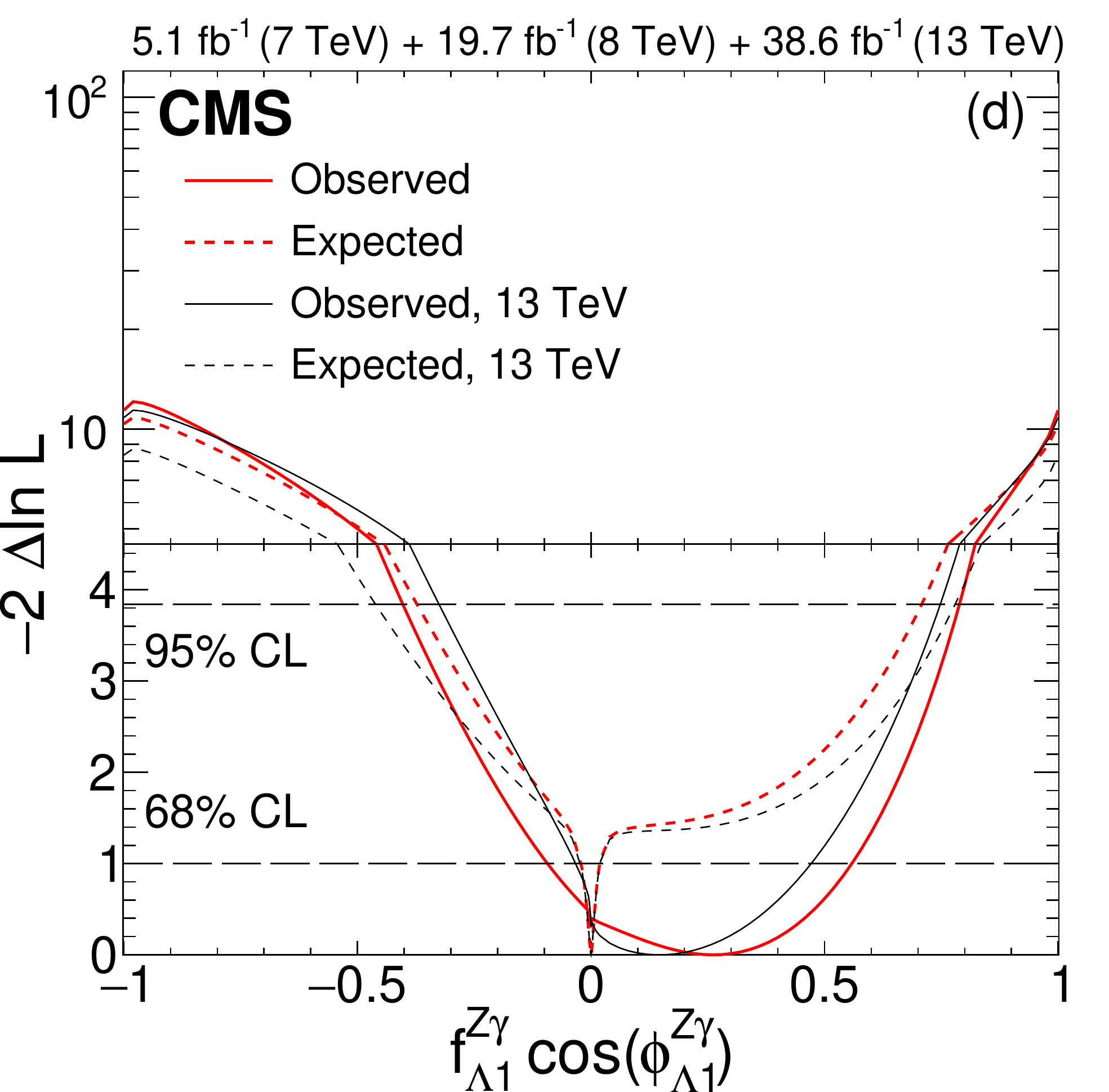}
\caption{
Observed (solid) and expected (dashed) likelihood scans of $f_{a3}\cos(\phi_{a3})$  (a),  $f_{a2}\cos(\phi_{a2})$ (b),
$f_{\Lambda1}\cos(\phi_{\Lambda1})$  (c), and $f_{\Lambda1}^{\Z\gamma}\cos(\phi_{\Lambda1}^{\Z\gamma})$ (d).
Results of the Run\,2 only and the combined Run\,1 and Run\,2 analyses are shown.
\label{fig:resultsfan}}
\end{figure*}

Four $f_{ai}$ parameters sensitive to anomalous $\PH$ boson interactions,
as defined in Eqs.~(\ref{eq:fai}) and~(\ref{eq:fai-inverse}),
are tested in the observed data using the pdf in Eq.~(\ref{eq:probability}).
The results of the likelihood scans of the $f_{ai}$ parameters on 13\TeV data only and on the full, combined data set
from collisions at 13, 8, and 7\TeV are shown in Fig.~\ref{fig:resultsfan}.
The combined results are listed in Table~\ref{tab:summary_spin0} and supersede our previous measurement
in Ref.~\cite{Khachatryan:2014kca}.

\begin{table*}
\centering
\topcaption{
Summary of allowed 68\%~\CL (central values with uncertainties) and 95\%~\CL (in square brackets)
intervals on anomalous coupling parameters obtained from the combined Run\,1 and Run\,2 data analysis.
}
\renewcommand{\arraystretch}{1.25}
\begin{tabular}{ccc}
\hline
 Parameter                                   &  {Observed} & {Expected}    \\
\hline
 $f_{a3}\cos(\phi_{a3})$         & $0.00^{+0.26}_{-0.09}$ $[-0.38,0.46]$ & $0.000^{+0.010}_{-0.010}$ $[-0.25,0.25]$  \\
 $f_{a2}\cos(\phi_{a2})$         & $0.01^{+0.12}_{-0.02}$ $[-0.04,0.43]$ & $0.000^{+0.009}_{-0.008}$ $[-0.06,0.19]$   \\
 $f_{\Lambda1}\cos(\phi_{\Lambda1})$  & $0.02^{+0.08}_{-0.06}$ $[-0.49,0.18]$ & $0.000^{+0.003}_{-0.002}$ $[-0.60,0.12]$  \\
 $f_{\Lambda1}^{\Z\gamma}\cos(\phi_{\Lambda1}^{\Z\gamma})$ & $0.26^{+0.30}_{-0.35}$ $[-0.40,0.79]$ & $0.000^{+0.019}_{-0.022}$ $[-0.37,0.71]$  \\
\hline
\end{tabular}
\label{tab:summary_spin0}
\end{table*}

The expected 68\%~\CL constraints improve by nearly an order of magnitude compared to the Run\,1
analysis~\cite{Khachatryan:2014kca}, as is evident from the narrow minima at $f_{ai}=0$ in the expectations
in Fig.~\ref{fig:resultsfan}. This effect comes from utilizing production information,
because the cross section in VBF and $\V\PH$ production increases quickly with $f_{ai}$
due to larger $q^2$ values contributing in Eq.~(\ref{eq:formfact-fullampl-spin0})~\cite{Anderson:2013afp}.
The narrow minima are shallower than expected, which may be understood by examining
the best fitted $(\mu_V, \mu_F$) values in the four analyses under the assumption that $f_{ai}=0$:
$(0.76^{+1.10}_{-0.76},1.08^{+0.21}_{-0.20})$ at $f_{a3}=0$,
$(0.01^{+0.89}_{-0.01},1.24^{+0.20}_{-0.18})$ at $f_{a2}=0$,
$(0.20^{+0.94}_{-0.20},1.20^{+0.21}_{-0.20})$ at $f_{\Lambda1}=0$,
and $(0.24^{+0.84}_{-0.24},1.20^{+0.20}_{-0.19})$ at $f_{\Lambda1}^{\Z\gamma}=0$.
The values obtained for the different analyses vary due to the different categorization
and observables.
The overall behavior with $\mu_V$ less than 1 is consistent with a downward statistical fluctuation
in the small number of VBF and $\V\PH$ events, with the total observed number of untagged events similar
to the expectation.
Because fewer VBF and $\V\PH$ events are observed than expected, the narrow minima of
$-2\ln(\mathcal{L})$ at $f_{ai}=0$,
which come from the production information in these events,
are observed to be less pronounced than expected.
The minimum is most pronounced in the $f_{a3}$ analysis in Fig.~\ref{fig:resultsfan} (a) due to the largest
observed $\mu_V$ value.

The improvement in the 95\%\,CL constraints with respect to Run\,1 is mostly due to the increase
in the number of events with $\PH\to4\ell$ decay information by about a factor of four.
Another factor of four increase in the data sample size is expected by the end of 2018,
under similar running conditions.  At that time, the inclusion of
production information is expected to result in improvements to the 95\% CL
constraints in line with the improvements already seen in the 68\% CL
constraints.

Other features in Fig.~\ref{fig:resultsfan} can be explained by examining the kinematic distributions in Fig.~\ref{fig:discriminants}.
The $\mathcal{D}_{0-}^\text{dec}$ distribution in Fig.~\ref{fig:discriminants}\,(e) favors a mixture
of the $f_{a3}=0$ and $f_{a3}=1$ models, resulting in the best fit value of $f_{a3}=0.30\pm0.21$ in Run\,2.
The $\mathcal{D}_{C\!P}^\text{dec}$ distribution in Fig.~\ref{fig:discriminants}\,(h) has a small forward-backward
asymmetry, with more events at $\mathcal{D}_{C\!P}^\text{dec}>0$ than $\mathcal{D}_{C\!P}^\text{dec}<0$,
which gives preference to the $f_{a3}\cos(\phi_{a3})=+0.30$ value as opposed to $-0.30$.
The narrow local minimum at $f_{a3}=0$ corresponds to the distribution of events in the tagged categories
in Fig.~\ref{fig:discriminants}\,(f),\,(g), which favors the SM hypothesis.
The Run 1 result [13] favors the SM strongly, and therefore combining
the two data sets results in a global minimum at $f_{a3}=0$.

Certain values of anomalous couplings, such as $f_{a2}\cos(\phi_{a2})\sim-0.5$ and
$f_{\Lambda1}\cos(\phi_{\Lambda1})\sim+0.5$, lead to strong interference effects between the SM
and anomalous amplitudes in Eq.~(\ref{eq:formfact-fullampl-spin0}).
Therefore, kinematic distributions of such models are easily distinguished from SM distributions,
and they are excluded at high \CL in Fig.~\ref{fig:resultsfan}.
Such anomalous models are shown in Fig.~\ref{fig:discriminants}\,(b),\,(c). The $f_{a3}=1$ and $f_{\Lambda1}^{\Z\gamma}=1$
models are shown in other cases in Fig.~\ref{fig:discriminants}, as the most distinct from SM,
except for (h), where maximal forward-backward asymmetry in $\mathcal{D}_{C\!P}$ is shown for $f_{a3}=0.5$.
In all cases, the observed distributions in Fig.~\ref{fig:discriminants} are consistent with the SM expectations.

\section{Summary}
\label{sec:Summary}

We study anomalous interactions of the $\PH$ boson using novel techniques with a
matrix element likelihood approach to simultaneously analyze
the $\PH\to4\ell$ decay and associated production with two quark jets.
Three categories of events are analyzed, targeting events produced
in vector boson fusion, with an associated vector boson, and in gluon fusion, respectively.
The data collected at a center-of-mass energy of 13\TeV in Run\,2 of the LHC
are combined with the Run\,1 data, collected at 7 and 8\TeV.
No deviations from the standard model are observed and constraints are set on the four anomalous $\PH\V\V$
contributions, including the $C\!P$-violation parameter $f_{a3}$, summarized in Table~\ref{tab:summary_spin0}.

\begin{acknowledgments}\label{sec:Acknowledgments}
We thank Markus Schulze for optimizing the \textsc{JHUGen} Monte Carlo simulation program and matrix element library for this analysis.
We congratulate our colleagues in the CERN accelerator departments for the excellent performance of the LHC and thank the technical and administrative staffs at CERN and at other CMS institutes for their contributions to the success of the CMS effort. In addition, we gratefully acknowledge the computing centers and personnel of the Worldwide LHC Computing Grid for delivering so effectively the computing infrastructure essential to our analyses. Finally, we acknowledge the enduring support for the construction and operation of the LHC and the CMS detector provided by the following funding agencies: BMWFW and FWF (Austria); FNRS and FWO (Belgium); CNPq, CAPES, FAPERJ, and FAPESP (Brazil); MES (Bulgaria); CERN; CAS, MoST, and NSFC (China); COLCIENCIAS (Colombia); MSES and CSF (Croatia); RPF (Cyprus); SENESCYT (Ecuador); MoER, ERC IUT, and ERDF (Estonia); Academy of Finland, MEC, and HIP (Finland); CEA and CNRS/IN2P3 (France); BMBF, DFG, and HGF (Germany); GSRT (Greece); OTKA and NIH (Hungary); DAE and DST (India); IPM (Iran); SFI (Ireland); INFN (Italy); MSIP and NRF (Republic of Korea); LAS (Lithuania); MOE and UM (Malaysia); BUAP, CINVESTAV, CONACYT, LNS, SEP, and UASLP-FAI (Mexico); MBIE (New Zealand); PAEC (Pakistan); MSHE and NSC (Poland); FCT (Portugal); JINR (Dubna); MON, RosAtom, RAS, RFBR and RAEP (Russia); MESTD (Serbia); SEIDI, CPAN, PCTI and FEDER (Spain); Swiss Funding Agencies (Switzerland); MST (Taipei); ThEPCenter, IPST, STAR, and NSTDA (Thailand); TUBITAK and TAEK (Turkey); NASU and SFFR (Ukraine); STFC (United Kingdom); DOE and NSF (USA).

\hyphenation{Rachada-pisek} Individuals have received support from the Marie-Curie program and the European Research Council and Horizon 2020 Grant, contract No. 675440 (European Union); the Leventis Foundation; the A. P. Sloan Foundation; the Alexander von Humboldt Foundation; the Belgian Federal Science Policy Office; the Fonds pour la Formation \`a la Recherche dans l'Industrie et dans l'Agriculture (FRIA-Belgium); the Agentschap voor Innovatie door Wetenschap en Technologie (IWT-Belgium); the Ministry of Education, Youth and Sports (MEYS) of the Czech Republic; the Council of Science and Industrial Research, India; the HOMING PLUS program of the Foundation for Polish Science, cofinanced from European Union, Regional Development Fund, the Mobility Plus program of the Ministry of Science and Higher Education, the National Science Center (Poland), contracts Harmonia 2014/14/M/ST2/00428, Opus 2014/13/B/ST2/02543, 2014/15/B/ST2/03998, and 2015/19/B/ST2/02861, Sonata-bis 2012/07/E/ST2/01406; the National Priorities Research Program by Qatar National Research Fund; the Programa Clar\'in-COFUND del Principado de Asturias; the Thalis and Aristeia programs cofinanced by EU-ESF and the Greek NSRF; the Rachadapisek Sompot Fund for Postdoctoral Fellowship, Chulalongkorn University and the Chulalongkorn Academic into Its 2nd Century Project Advancement Project (Thailand); and the Welch Foundation, contract C-1845.
\end{acknowledgments}

\bibliography{auto_generated}

\cleardoublepage \appendix\section{The CMS Collaboration \label{app:collab}}\begin{sloppypar}\hyphenpenalty=5000\widowpenalty=500\clubpenalty=5000\textbf{Yerevan Physics Institute,  Yerevan,  Armenia}\\*[0pt]
A.M.~Sirunyan, A.~Tumasyan
\vskip\cmsinstskip
\textbf{Institut f\"{u}r Hochenergiephysik,  Wien,  Austria}\\*[0pt]
W.~Adam, F.~Ambrogi, E.~Asilar, T.~Bergauer, J.~Brandstetter, E.~Brondolin, M.~Dragicevic, J.~Er\"{o}, M.~Flechl, M.~Friedl, R.~Fr\"{u}hwirth\cmsAuthorMark{1}, V.M.~Ghete, J.~Grossmann, J.~Hrubec, M.~Jeitler\cmsAuthorMark{1}, A.~K\"{o}nig, N.~Krammer, I.~Kr\"{a}tschmer, D.~Liko, T.~Madlener, I.~Mikulec, E.~Pree, D.~Rabady, N.~Rad, H.~Rohringer, J.~Schieck\cmsAuthorMark{1}, R.~Sch\"{o}fbeck, M.~Spanring, D.~Spitzbart, W.~Waltenberger, J.~Wittmann, C.-E.~Wulz\cmsAuthorMark{1}, M.~Zarucki
% ...
\vskip\cmsinstskip
% ...
1:~~Also at Vienna University of Technology, Vienna, Austria\\
% ...

\end{sloppypar}
\end{document}